 \documentclass[preprint2,longabstract]{aastex}

\shorttitle{A Nuclear Data Approach for the Hubble Constant Measurements}
\shortauthors{B. Pritychenko}

\begin{document}


\title{A Nuclear Data  Approach for the Hubble Constant Measurements}


\author{B. Pritychenko}
\affil{National Nuclear Data Center, Brookhaven National Laboratory, Upton, NY 11973-5000}


%\altaffiltext{1}{Visiting Astronomer, Cerro Tololo Inter-American Observatory.
%CTIO is operated by AURA, Inc.\ under contract to the National Science


\begin{abstract}
An extraordinary number of Hubble constant measurements challenges physicists with selection of  the best numerical value. 
The standard U.S. Nuclear Data Program (USNDP) codes and procedures have been applied to resolve this issue. 
The nuclear data approach  has produced the most probable or recommended Hubble constant value of 66.2(77)   (km/sec)/Mpc. 
This  recommended value is based on the last 25 years of experimental research and includes contributions from different 
types of measurements. The present result  implies   (14.78$\pm$1.72)$\times$10$^{9}$ years as a rough estimate for the age of the Universe. 
The complete list of recommended results is given and possible implications are discussed.
\end{abstract}


\keywords{cosmology: observations ---  methods: data analysis}



\section{Introduction}

\begin{figure*}
\includegraphics[angle=0,scale=.50]{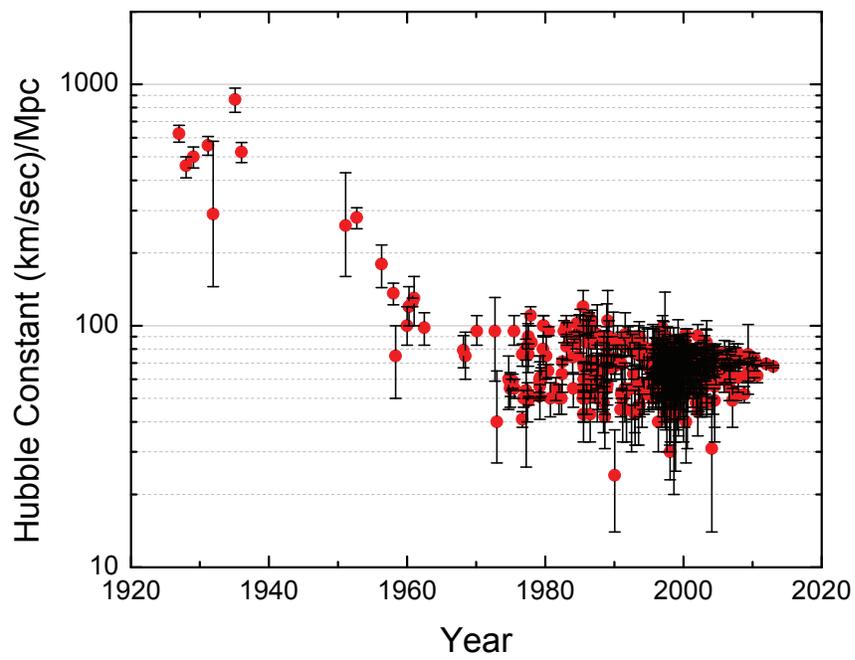}
\caption{Historical evolution of Hubble constant measurements. 
%Plots for all sources are available in the electronic edition of {\it The Astrophysical Journal}.
\label{fig1}}
\end{figure*}

The Hubble constant and its present-day numerical value play an important role in modern astrophysics and 
cosmology \cite[]{boyd2008,dolgov1988}. For many years,  researchers have been improving the accuracy of the constant \cite[]{hubble2013}. 
Fig. ~\ref{fig1} shows  the time evolution of Hubble research in the last 100 years, and the original effort can be 
traced as early as 1916 \cite[]{huchra2010}. The large number of measurements creates a certain degree of confusion 
about Hubble constant numerical value, and scientists often rely on  recently-published results \cite[]{pdg2014}. 
The precision of  Hubble constant measurements has improved enormously over the years; however, it is not always prudent 
to reject older results in favor of the latest findings.  Consequently, it makes perfect sense to analyze all available results, 
evaluate the data,  and extract the  recommended  value.

\section{Hubble Constant Evaluation}

The volume of Hubble constant measurements  far exceeds other experimental quantities in physics \cite[]{pritychenko2015}.  
Previously, similar situations have been encountered in nuclear and particle physics and resolved with data evaluations. 
 Nuclear data evaluations and their policies are well described in literature \cite[]{pritychenko2012,ensdf2015}.  Frequently, 
 the evaluations are completely based on or adjusted to  available experimental data.   
 These adjustments and specialized mathematical statistics techniques  can be applied for nuclear, particle,  or any other data sets.
  
 In this work, I would follow standard nuclear data evaluation procedures to deduce the recommended value.   
 Current evaluation input  data   are mostly based on the  NASA/HST Key Project on the Extragalactic Distance Scale 
 compilation \cite[]{huchra2010}   and recent results.  
  A visual inspection   of historical Hubble Constant measurements, as shown in  Fig.~\ref{fig1},  is instrumental in the data analysis.   
  It suggests  that one may safely reject all  measurements prior to 1970.  It is common knowledge that Hubble constant measurements 
  heavily rely on the accuracy of astronomic distance determination. Older results, such as those reported by 
  A. Sandage and  G. de Vancouleurs, suffered from  inaccurate measurements  \cite[]{hubble2013}. Therefore, the rejection of 
  all results prior to 1990 could provide a complimentary benchmark value of the Hubble constant.

 In the present data analysis, the experimental  data have been separated into two groups, with 1970 and 1990 time cuts, 
 and further reduced using the following policies:
\begin{itemize}
\item Rejection of repeated results (multiple publication of the same result)
\item Rejection of model-dependent results (i.e. Cosmic Microwave Background (CMB) fits)
\item Rejection of potential outliers using Chauvenet's criterion \cite[]{birch2014}
\end{itemize}

Common data evaluation practices indicate that recommended value should be based on a large statistical sample 
that includes different types of measurements. The 1970 and 1990 redacted data sets   of $\sim$334 and $\sim$266 data points, respectively, 
provide such samples. These large samples create the possibility of deducing Hubble constant value for each method of 
observation besides the combined value that is based on all measurements.    The current data collections were further 
subdivided using a NASA/HST Key Project on the Extragalactic Distance Scale classification of experimental methods \cite[]{huchra2010}: 
\begin{itemize}
\item S = Type Ia supernovae (SNIa)
\item 2 = Type II supernovae (SNII)
%\item G = Globular cluster luminosity function
\item L = Lens         
\item r = Red Giants
\item B = Baryonic Tully-Fisher
\item R = Inverse Tully-Fisher (ITF,RTF)
\item H = Infrared Tully-Fisher (or IRTF)
\item F = Fluctuations 
%\item D = Dn-Sig/Fund Plane
\item A = Global Summary     
\item Z = Sunyaev-Zeldovich      
\item T = Tully-Fisher
\item O = Other  
%\item N = Novae   
%\item P = Planetary nebula luminosity function                   
%\item C = CMB Fit
\end{itemize}

These experimental data sets have been processed with the latest version of the visual averaging library \cite[]{birch2014}. 
The library program includes limitation of relative statistical weight (LWM), normalized residual (NRM), Rajeval technique (RT), 
and the Expected Value (EVM) statistical methods to calculate averages of experimental data with uncertainties. 
The experimental  data sets were processed, and evaluated values with reduced $\chi^2$$< $2 were typically accepted as reasonable data fits. 
The current evaluation incorporates statistical methods based on the inverse squared value of the quoted uncertainties, 
a  procedure that is consistent with the general methodology used in treatment of data for the ENSDF database \cite[]{ensdf2015} 
and Particle Data Group  \cite[]{pdg2014}.

\section{Results and Discussion}

\begin{figure*}
\includegraphics[angle=0,scale=.50]{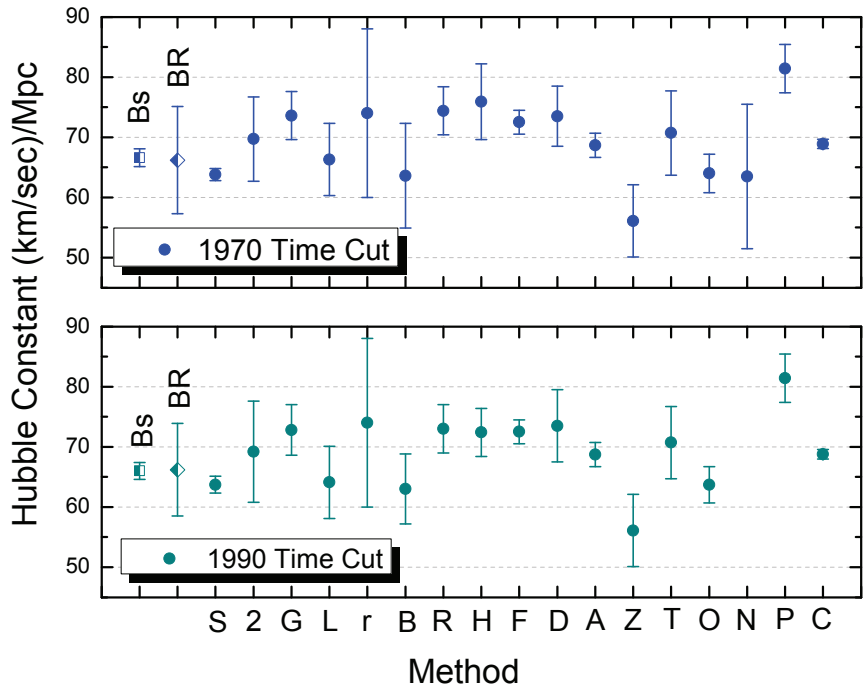}
\caption{Evaluated values of Hubble constant  using  1970 and 1990 time cuts for experimental data. These plots include the evaluated or recommended values for combined  observations, 
using the Best Representation (BR) and Bootstrap (Bs) data analysis techniques. % and recommendations for contributing  experimental methods that are described in the article. 
Globular cluster luminosity function (G), Dn-Sig/Fund Plane (D), Novae (N), Planetary nebula luminosity function (P)  results are also included while  CMB Fit (C)  value is shown for comparative purposes. 
%Plots for all sources are available in the electronic edition of {\it The Astrophysical Journal}.
\label{fig2}}
\end{figure*}

Two sets of  recommended values are displayed in Fig.~\ref{fig2}, and  the combined numerical values  are shown in Table~\ref{tbl1}.  
The Hubble constant combined central values extracted by means of different mathematical techniques are in good agreement, 
while uncertainties need further discussion. Visual inspection of the numerical values shown in the  Fig. ~\ref{fig2}  indicates the 
Best Representation mathematical procedure (depicted as ``BR") provides a best fit for different experimental methods 
with reasonable uncertainties. Other procedures  such as the Bootstrap technique (depicted as ``Bs") result in rather small uncertainties. 
These small uncertainties are due to specifics of data analysis procedures such as assignment of different statistical weights 
to results with different uncertainties.  In light of this disclosure, it is prudent to select the Best Representation results as 
the recommended value.  The current  results are consistent with the recent Particle Data Group publication \cite[]{pdg2014}. 

Finally, two different time cuts of 1970 and 1990 for Hubble's data have yielded two recommended values 
of 66.2(89) and  66.2(77)   (km/sec)/Mpc, respectively. The agreement between these values partially reflects the fact 
that the majority of the Hubble's constant measurements has been performed in the last 25 years, and a small number 
of potential outliers has been rejected. More accurate  recent observations imply a preference for the 1990 time cut 
value  of  66.2(77)   (km/sec)/Mpc. The last result is consistent with the Hubble Space Telescope and Wide Field Camera 3 \cite[]{riess2011} 
and  the model-dependent Planck's Mission and WMAP values \cite[]{Bucher2013,Bennett2013}. Inclusion  of the globular cluster luminosity function, 
Dn-Sig/Fund Plane, Novae and  planetary nebula luminosity function data would slightly change 1970 and 1990 
recommended values to 66.9(90) and 66.9(78), respectively. These less precise measurements do not affect the recommended values severely 
because of a  Chauvenet's criterion analysis.

In recent years, an effort has been made to calculate the Hubble constant median statistics \cite[]{chen2011}. The median statistics 
approach differs substantially from the nuclear or particle data evaluation procedures. At the same time, it provides  a complementary 
value of 68$\pm$5.5 (or $\pm$1) (km/sec)/Mpc \cite[]{chen2011} that can be compared with the present results.

The  knowledge of Hubble constant value has multiple implications in science. As an example, a rough estimate of the 
age of the Universe can be deduced using the standard methodic \cite[]{wiki2015}. The adopted value of 66.2(77) (km/sec)/Mpc implies  
(14.78$\pm$1.72)$\times$10$^{9}$ years estimated value for the age of Universe. The last result is consistent with  the recently 
published  value  of (13.798$\pm$0.037)$\times$10$^{9}$ years \cite[]{age2014}.

\section{Conclusions}
The analysis of Hubble constant measurements has been performed using standard USNDP codes and procedures. 
An evaluated data set of most probable values of Hubble constant has been  deduced and shown in the Table~\ref{tbl1}. 
These values are consistent with other available results. An accurate constant value is instrumental for many potential applications. 
The recommended value of the constant is completely based on experimental measurements, and further, more precise observations, 
would lead to more accurate determination of it.

\acknowledgments

The author is indebted to Dr. M. Herman (BNL) for support of this project  and grateful to Dr. V. Unferth (Viterbo University) 
for help with the manuscript. This work was funded by the Office of Nuclear Physics, Office of Science of the U.S. Department 
of Energy, under Contract No. DE-AC02-98CH10886 with Brookhaven Science Associates, LC.


%% in parentheses, after the keyword, but they will not be verified.

%{\it Facilities:} \facility{Nickel}, \facility{HST (STIS)}, \facility{CXO (ASIS)}.



\clearpage

%% the file, uncomment the \plotone, \plottwo, and
%% If you need a layout that cannot be achieved with \plotone or
%% \plottwo, you can invoke the graphicx package directly with the
%% \includegraphics command or use \plotfiddle. For more information,
%% please see the tutorial on "Using Electronic Art with AASTeX" in the
%% documentation section at the AASTeX Web site, http://aastex.aas.org/
%%
%% The examples below also include sample markup for submission of
%% supplemental electronic materials. As always, be sure to check
%% the instructions to authors for the journal you are submitting to
%% for specific submissions guidelines as they vary from

%% This example uses \plotone to include an EPS file scaled to
%% 80% of its natural size with \epsscale. Its caption
%% has been written to indicate that additional figure parts will be
%% available in the electronic journal.

\clearpage

%% Here we use \plottwo to present two versions of the same figure,
%% one in black and white for print the other in RGB color
%% for online presentation. Note that the caption indicates
%% that a color version of the figure will be available online.
%%

\clearpage

\begin{table*}
\begin{center}
\caption{Results of the Hubble Constant evaluation for all observations using 1970 and 1990 time cuts. \label{tbl1}}

\begin{tabular}{c|c|c}
\tableline\tableline
Method &	Time Cut: 1970	 & Time Cut: 1990 \\
              & (km/sec)/Mpc & (km/sec)/Mpc \\
\tableline
Unweighted Average &	67.56(73) &  65.92(65) \\
Weighted Average & 63.68(58)	 &  63.87(51) \\
LWM &	67.3(93)	&  66.08(64) \\
Normalized Residual & 64.50(50) & 64.33(48)  \\
Rajeval Technique & 65.57(45)	 & 65.07(46)  \\
Best Representation (BR) &	66.2(89)	 & 66.2(77)  \\
Bootstrap (BS) & 	66.6(15)	& 66.0.(14)  \\
Mandel-Paule &	66.7(92)	&  65.3(62) \\
\tableline\tableline
\end{tabular}
\end{center}
\end{table*}

\end{document}